\documentclass[manuscript]{aastex}
\usepackage{graphicx,natbib,amsmath,amssymb}
\begin{document}
\title{Dissipation Efficiency in Turbulent Convective Zones in Low Mass Stars}
\author{K. Penev and D. Sasselov}
\affil{Department of Astronomy, Harvard University, 
	60 Garden Street, Cambridge, MA 02138}
\author{F. Robinson}
\affil{Department of Geology and Geophysics, Yale University
	New Haven, CT 06520-8101}
\and
\author{P. Demarque}
\affil{Department of Astronomy, Yale University, New Haven, CT 06520-8101}

\begin{abstract}
We extend the analysis of \citet{Penev_Sasselov_Robinson_Demarque_07} to
calculate effective viscosities for the surface convective zones of three main
sequence stars of $0.775M_\odot$, $0.85M_\odot$ and the present day Sun. In
addition we also pay careful attention to all normalization factors and
assumptions in order to derive actual numerical prescriptions for the effective
viscosity as a function of the period and direction of the external shear. Our
results are applicable for
periods that are too long to correspond to eddies that fall within the inertial
subrange of Kolmogorov scaling, but no larger than the convective turnover
time, when the assumptions of the calculation break down. 

We find a significantly anisotropic viscosity, scaling linearly with the period
of the external perturbation and magnitudes of the different components between
three and ten times smaller than the \citet{Zahn_66, Zahn_89} prescription.
\end{abstract}
\section{Introduction}
	Turbulent (eddy) viscosity
        is often considered to be the main mechanism responsible for dissipation
	of tides and oscillations  in convection zones of cool stars and
	planets (\citet{Goodman_Oh_97}(from now on GO), and references therein).
	Currently existing descriptions have been used, with varying
	success, to explain circularization cut-off periods for main
	sequence binary stars \citep{Zahn_Bouchet_89, Meibom_Mathieu_05}, 
	the red edge of the Cepheid instability strip \citep{Gonczi_82}
	and damping of solar oscillations \citep{Goldreich_Keeley_77}.
        However, this hypothesis has been far more successful in damping 
	oscillations than damping tides, and different mechanisms have been 
	proposed for the latter, especially for planets (see \citet{Wu_05a,
	Wu_05b, Ogilvie_Lin_04} and references therein).
		
        The standard treatment is to  assume
	a Kolmogorov spectrum in the convection
	zone and apply some prescription to model the effectiveness of  eddies
	in dissipating the given perturbation. Two prescriptions have
	been proposed to describe the efficiency of eddies in  dissipating
	perturbations with periods ($T$) smaller than the eddy turnover
	time ($\tau$).

	The first prescription, due to \citet{Zahn_66, Zahn_89}, assumes that
	it is always the largest eddies that dominate the dissipation and that
	they lose efficiency linearly with decreasing period:
	\begin{equation}
		\nu = \nu_{max} \min\left[ 
			\left(\frac{T}{2\tau_c}\right),1\right]
		\label{eq: Zahn viscosity}
	\end{equation}
	Where $\nu_{max}$ is some constant which depends on  the mixing length
	parameter and $\tau_c$ is the local convective turnover time (or the
	turnover time of the largest local eddies). 
	This prescription has been tested against  tidal
	circularization times for binaries containing a giant star
	\citep{Verbunt_Phinney_95}, and is in general agreement with
	observations.\\

	The second prescription, due to \citet{Goldreich_Nicholson_77} and 
	\citet{Goldreich_Keeley_77}, assumes that eddies with periods longer
	than $T/2\pi$ do not contribute to the dissipation. In that case, for
	Kolmogorov scaling:
	\begin{equation}
		\nu = \nu_{max} \min\left[
				\left(\frac{T}{2\pi\tau_c}\right)^2,1\right]
	\end{equation}

	This prescription has been used successfully by 
	\citet{Goldreich_Keeley_77, Goldreich_Kumar_88} and 
	\citet{Goldreich_Kumar_Murray_94} to develop a theory for the damping of the 
	solar $p$-modes. If the more effective dissipation was applied
	instead, dramatic changes would be required in the excitation
	mechanism in order to explain the observed  $p$ mode
	amplitudes. However, this inefficient dissipation is
	inconsistent with observed tidal circularization for binary
	stars \citep{Meibom_Mathieu_05}. Additionally, \citet{Gonczi_82}
	argues
	that for pulsating stars the location of the red edge of the
	instability strip is more consistent with Zahn's description
	of eddy viscosity than with that of Goldreich and
	collaborators.\\

	GO gave a consistent hydrostatic
	derivation of the convective viscosity, using a perturbation
	approach. For a Kolmogorov scaling they obtained a result that is
	closer to the less efficient Goldreich \& Nicholson viscosity than it is
	to Zahn's, providing a more sound theoretical
	basis for the former scaling. Of course, the observational problem of
	insufficient tidal dissipation for stellar pulsations and binaries
	remains unresolved, as GO point out.\\

	Both 2D and 3D numerical simulations of the solar convection
	zone have revealed that the picture of a Kolmogorov spectrum
	of eddies is too simplified 
	\citep{Stein_Nordlund_89, Robinson_et_al_03}. The  simulations showed
	that convection proceeds in a rather different,  highly
	asymmetric fashion. This suggests that the problem of
	insufficient dissipation may be resolved by replacing the
	assumption of Kolmogorov turbulence with the velocity
	field produced from numerical simulations. More importantly,
	an asymmetric and non-Kolmogorov turbulence might dissipate
	different perturbations differently, i.e. depending both on
	the frequency and geometry of the perturbation. Such 
	simulations have been used to develop a better model for the 
	excitation of solar $p$-modes \citep{Samadi_et_al_03}.\\

	In \citet{Penev_Sasselov_Robinson_Demarque_07} we 
	reconsidered the problem of tidal dissipation 
	in stellar convection zones of solar-type stars by applying the
	approximation developed in GO to the turbulent 
	velocity field from a realistic 3D solar simulation and showed that,
	the scaling predicted by this procedure is very close to linear.
	The shallower scaling is explained by the fact that on the time scales 
	captured by the simulation the largest eddies have typical sizes 
	comparable to the local pressure scale height and, in this regime,
	the velocity power spectrum is much shallower than Kolmogorov.

	In this paper we apply a more complete version of the same scheme 
	to three stellar convection models appropriate for stars with
	masses of $0.775M_\odot$, $0.85M_\odot$ and $1M_\odot$. We also pay
	significantly more attention to the normalization and the approximations
	which we introduce in addition to GO.
\section{Method}
\subsection{The Perturbative Expansion for Discretely Sampled Velocity Field}
We follow the procedure outlined in GO and assume an external perturbing
velocity field given by GO equation (8):
\begin{equation}
	\mathbf{V}_t=\mathbf{A}(t)\cdot\mathbf{x},
\end{equation}
where $\mathbf{A}(t)$ is some matrix that depends only on time, and not
space. This is appropriate when the spatial dependence of the
perturbation is on scales much larger than our simulation box (e.g.
tides). The matrix $\mathbf{A}(t)$ is assumed symmetric, since the
antisymmetric part corresponds to rotation, and is not expected to contribute
to the energy dissipation.

Introducing this velocity field will modify the convective flow 
($\mathbf{v_0}$). The time evolution of the change in the
turbulent velocity ($\delta \mathbf{v}$) due to the presence of the
above external field can be written as in GO, equation (19):
\begin{equation}
	\partial_t \delta \mathbf{v}(\mathbf{x},t) = -2\mathbf{A}(t)
	\cdot \mathbf{v_0}(\mathbf{x}, t) -
	\mathbf{v}_0\cdot\mathbf{\nabla}\delta \mathbf{v} -
	\delta\mathbf{v}\cdot\mathbf{\nabla}\mathbf{v_0} - 
	(\textrm{pressure term}).
	\label{eq: delta v phys}
\end{equation}

GO assumed incompressibility, and hence the pressure term simply
maintains that $\mathbf{\nabla}\cdot\delta\mathbf{v}=0$. We use the
output of fully compressible simulations, so for us the pressure term
should be much more complex. However, the only type of compressibility that we
can reasonably incorporate in the analysis is that due to the stratification of
the convective layer, and even that we approximate by assuming a constant
density scale height. This is discussed in more detail in
section \ref{sec: compressibility}.

We then follow GO in writing equation (\ref{eq: delta v phys}) in Fourier
space. However, since we are dealing with discretely sampled data, 
we use discrete Fourier transforms:
\begin{eqnarray}
	\delta\mathbf{v}_{l,m,n,p}&=&\frac{1}{N_x N_y N_z N_t}
	\sum_{\lambda, \mu, \nu, \phi} \widehat{\delta\mathbf{v}}_{\lambda,
	\mu, \nu, \phi} e^{2\pi i \left(
		\frac{\lambda l}{N_x} + \frac{\mu m}{N_y} +
		\frac{\nu n}{N_z} + \frac{\phi p}{N_t}
	\right)}\nonumber,\\
	{\mathbf{v}_0}_{l,m,n,p}&=&\frac{1}{N_x N_y N_z N_t}
	\sum_{\lambda, \mu, \nu, \phi} \widehat{\mathbf{v}}_{0\lambda,
	\mu, \nu, \phi} e^{2\pi i \left(
		\frac{\lambda l}{N_x} + \frac{\mu m}{N_y} +
		\frac{\nu n}{N_z} + \frac{\phi p}{N_t}
	\right)},\\
	\mathbf{A}(t)&=&\frac{1}{2}\left[\widehat{\mathbf{A}}(\Omega) 
	e^{-i\Omega t}+\widehat{\mathbf{A}}(-\Omega)e^{i\Omega t}
	\right]\nonumber,
\end{eqnarray}
where $2\pi/\Omega$ is the period of the external forcing.

For more details on how exactly the Fourier transform is applied in the
radial and time directions see section. \ref{sec: Fourier windows}. 

In Fourier space to first order in $\mathbf{A}$ (the strength of the
perturbation) and $\Omega\tau_c$ (the ratio of perturbation time scale
to convective turnover time) equation (\ref{eq: delta v phys}) is written as:
\begin{equation}
	\delta \hat{v}_{\lambda,\mu,\nu,\phi} = -\frac{i}{\omega_\phi}
	\mathbf{P}_{\lambda,\mu,\nu} \left[
	\mathbf{\widehat{A}}(\Omega)\cdot\mathbf{\hat{v}}(\omega_\phi-\Omega, 
		\mathbf{k}_{\lambda,\mu,\nu})
	+\mathbf{\widehat{A}}(-\Omega)\cdot\mathbf{\hat{v}}(\omega_\phi+\Omega, 
		\mathbf{k}_{\lambda,\mu,\nu})
	\right],
	\label{eq: v perturbation}
\end{equation}
where $P_{\lambda,\mu,\nu}\equiv
\mathbf{I}-
\mathbf{k'}_{\lambda,\mu,\nu} \mathbf{k'}_{\lambda,\mu,\nu}/
k'^2_{\lambda,\mu,\nu}$, with $\mathbf{k'}_{\lambda,\mu,\nu}\equiv
k_{\lambda,\mu,\nu}+i\hat{z}/H_\rho$, is the 
discrete version of the projection
operator GO define that imposes compressibility due only to a constant density
scale height.

We can then express the average rate of work done (per unit volume) on
the turbulent velocities by the tide to lowest non-zero order as:
\begin{equation}
	\begin{array}{c}
	\displaystyle{
	S_{\rho,\rho'}\equiv\frac{T}{\mathcal{N}^2 N_z} 
	\sum_{\lambda,\mu,\nu,\nu'}
	\rho_{\nu-\nu'}^* 
	v^1_{\lambda,\mu,\nu,\rho} P_{\lambda,\mu,\nu'} 
	v^{2*}_{\lambda,\mu,\nu',\rho'}}\\
	\displaystyle{\dot{\mathcal{E}}(\Omega=2\pi R/T)
	= \textrm{Re}\left\{S_{R,-R}+S_{R,R}\right\}
	+ \sum_{r\neq0} \dfrac{1}{\pi r}
	\textrm{Im}\left\{S_{r+R, r-R} + S_{r+R, r+R}\right\}},
	\end{array}
	\label{eq: e dot}
\end{equation}
where $\mathcal{N}\equiv N_x N_y N_z N_t$. 
In keeping with GO, we have assumed that all frequencies have an
infinitesimal imaginary part, which gives rise to the first term above.
The second term is entirely due to the density stratification in the
box: in the case of $\rho(z)=const$ it is zero. This term is the most important
difference between this calculation and GO.

In deriving equation
(\ref{eq: e dot}) we have assumed that the density is only a function of
depth. Keeping the radial dependence is necessary because the simulation 
box encompasses several density
scale heights, while the horizontal and temporal dependence of the
density is a much smaller effect, entirely due to the 
turbulent fluctuations in the box and if we could average over different 
realizations of the turbulence they would not be present. 

\subsection{Anisotropic Viscosity}
\label{sec: anisotropic_visc}
In order to extract an effective viscosity, we need to express the energy
dissipation rate that would occur in the presence of actual anisotropic
viscosity. 

Most generally, viscosity is a fourth order tensor relating the
strain, given by $\mathbf{A}(t)$ in this case, and the viscous stress:
\begin{equation}
	\sigma_{ij}^{visc} = K_{ijmn} A_{mn}(t).
\end{equation}

With this definition the time averaged dissipated power is given by:
\begin{equation}
	\dot{\mathcal{E}}_{visc}(\Omega)=\frac{1}{2}\int_0^{L_z} dz K_{ijmn}(z)
	Re\left\{A_{ij}(\Omega)A^*_{mn}(\Omega)\right\},
	\label{eq: viscous power}
\end{equation}
where, in order to remain consistent with the Fourier transform
conventions we simply replace the integral with a sum. To get the different
components of $K_{ijmn}$ we evaluate equation (\ref{eq: e dot}) with
$\mathbf{A}$ having nonzero elements at different locations, and use the above
equation to find the respective viscosity coefficients.

The viscosity tensor ($K_{ijmn}$) obeys a set of symmetries
that dramatically reduce the number of independent components. Since the strain
rate is symmetric by definition, and the stress must be symmetric in order to
keep the viscous torque on infinitesimal fluid elements finite, we must have:
\begin{equation}
	K_{ijmn}=K_{jimn}=K_{ijnm}.
\end{equation}
In addition, the only distinct direction in the problem is that of gravity
($\hat{z}$), so we expect the viscosity tensor to  be symmetric with respect to
rotation around the vertical axis. 

With all these symmetries we are left with only six independent components of
$K_{ijmn}$: $K_{1111}$, $K_{3333}$, $K_{1212}$, $K_{1313}$, $K_{1133}$ and
$K_{3311}$. Since the last two of these always appear together in the expression
for the energy dissipation we will assume them to be equal. The remaining
non-zero components can be found from those as follows:
\begin{equation}
	\begin{array}{r@{=}l}
	K_{2222}		&	K_{1111}, \\
	K_{1122}=K_{2211}	&	K_{1111}-2K_{1212}, \\
	K_{1221}=K_{2121}=K_{2112}&	K_{1212}, \\
	\left.\begin{array}{r}
	K_{3131}=K_{3113}=K_{1331}=K_{2323}\\
	K_{3232}=K_{3223}=K_{2332}
	\end{array}\right\}	&	K_{1313},\\
	K_{2233}&K_{1133},\\
	K_{3322}&K_{3311}.
	\end{array}
\end{equation}

A more physically meaningful set of five viscosity components can be found by
noting that under these symmetries the strain rate has only four distinct
components:
\begin{equation}
	\mathcal{A}_0\equiv A_{11}+A_{22};\quad 
	\mathcal{A}_{0'}\equiv A_{33};\quad 
	\mathcal{A}_{1}\equiv A_{13}+iA_{23};\quad
	\mathcal{A}_{2}\equiv A_{11}-A_{22}+2iA_{12},
	\label{eq: rotation A}
\end{equation}
along with their complex conjugates $\mathcal{A}_{-m}=\mathcal{A}_{m}^*$,
which transform under rotation by angle $\theta$ around the $\hat{z}$ axis as
$\mathcal{A}_m\rightarrow e^{i\theta m}\mathcal{A}_{m}$.

Clearly then, if $\dot{\mathcal{E}}_{visc}(\Omega)$ is to be invariant under
such rotations, it must be of the form:
\begin{equation}
	\dot{\mathcal{E}}_{visc}(\Omega)=\frac{1}{2}\int_0^{L_z} dz \left[
	4K_1|\mathcal{A}_1|^2 + K_2|\mathcal{A}_2|^2 + K_0 \mathcal{A}_0^2 +
	K_{0'} \mathcal{A}_{0'}^2 + 2K_{00'} \mathcal{A}_0\mathcal{A}_{0'}
	\right],
	\label{eq: viscous power rot}
\end{equation}
where the five new viscosity coefficients can be expressed in terms of
$K_{ijmn}$ as follows:
\begin{equation}
	\begin{array}{rcl}
		K_0&=&\dfrac{1}{2}\left(K_{1111}+K_{1122}\right)\\
		K_{0'}&=&K_{3333}\\
		K_{00'}&=&K_{1133}\\
		K_1&=&K_{1313}\\
		K_2&=&\dfrac{1}{2}\left(K_{1111}-K_{1122}\right)\\
	\end{array}
	\label{eq: K relations}
\end{equation}

Since in equation (\ref{eq: viscous power}) we allow the viscosity to
depend on depth, and there is no way to constrain this dependence, we have to
choose some radial profile a priori. Our choice is motivated by mixing length
theory:
\begin{equation}
	K_m(z)=K_m^0(\Omega) \rho\left<v^2\right>^{1/2} H_p,
	\label{eq: nu form}
\end{equation}
where $K_m^0$ are dimensionless constants, that depend on the
frequency of the external shear ($\Omega$), and $H_p$ is the local pressure
scale height. This is reasonable, since the turbulent viscosity should
scale as some length scale times some velocity scale. Clearly the
relevant velocity scale is that of convection, and in accordance with
mixing length theory, we use the mixing length as the length scale, which is
assumed proportional to the pressure scale height. If
the mixing length is really the relevant quantity, we expect that the
value of $K_m^0$ will be proportional to the mixing length parameter for
the particular simulation. This same scaling has been assumed for all previous
effective viscosity prescriptions \citep{Zahn_66, Zahn_89, Goldreich_Keeley_77,
Goldreich_Kumar_88, Goldreich_Kumar_Murray_94}.

\subsection{The Pressure Term}
\label{sec: compressibility}
In deriving the expression for $\dot{\mathcal{E}}$ (Eq. \ref{eq: e dot}) 
we assumed that the perturbation to the convective velocity field due to the
tide will be anelastic: $\mathbf{\nabla}\cdot\delta\mathbf{v}+v_z/H_\rho=0$, with
$H_\rho=const$. In this section we define two diagnostics which measure how
important the ignored compressibility is.

There are two sources of compressibility in the convective flow: 
\begin{enumerate}
	\item The convective flow carrying parcels of matter through layers of
		different hydrostatic pressure, or in other words due to the
		stratification.
	\item Localized compression due to a possibly supersonic flow, e.g.
		shocks.
\end{enumerate}
Ignoring the second one is justified, as long as the flow velocity is much less
than the local speed of sound. In the simulations we use, that condition is met
by the unperturbed flow for most of the box, with the exception of the
supersonic driving region near the top. If the unperturbed flow is subsonic and
hence incompressible, the perturbations due to a ``small'' external field can
safely be assumed incompressible as well. To measure the compressibility in the
simulation box we introduce the parameter:
\begin{equation}
	\xi\equiv\tau_c\left[\mathbf{\nabla}+\hat{z}\frac{d\ln\rho}{dz}\right]
		\cdot\mathbf{v_0},
	\label{eq: compressibility}
\end{equation}
Where, $\rho$ is the density averaged over horizontal slices and time.

This quantity deviates from zero due
to localized, transient compressions (e.g. shocks). Since those are
unlikely to live longer than a convective turnover time, this quantity is a
suitable diagnostic for the importance of such effects.

\begin{figure}[t]
\begin{center}
	\includegraphics[width=0.49\textwidth]{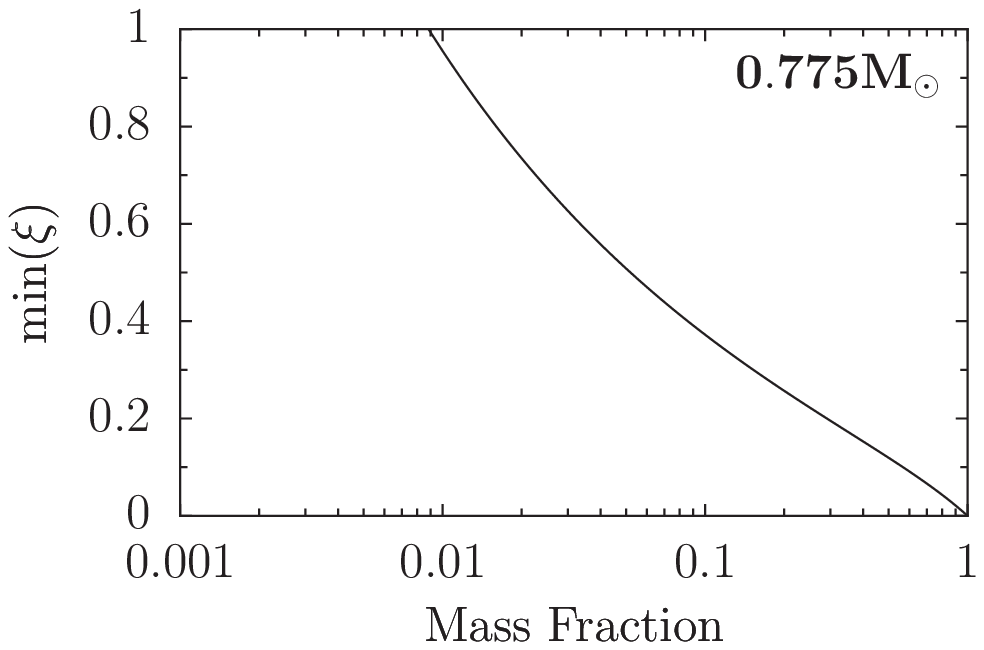}
	\includegraphics[width=0.49\textwidth]{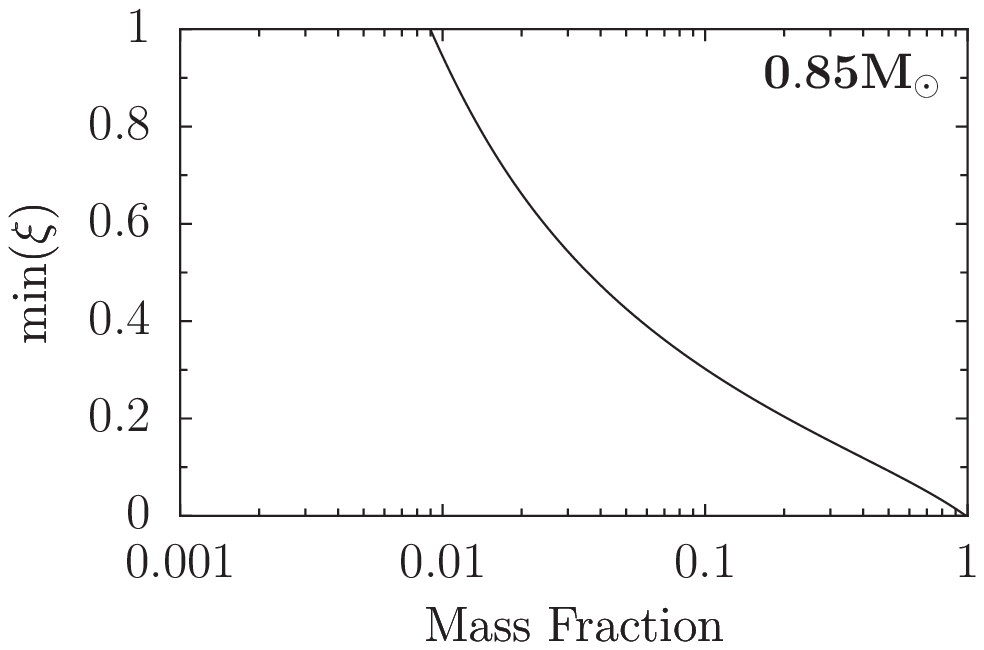}
	\includegraphics[width=0.49\textwidth]{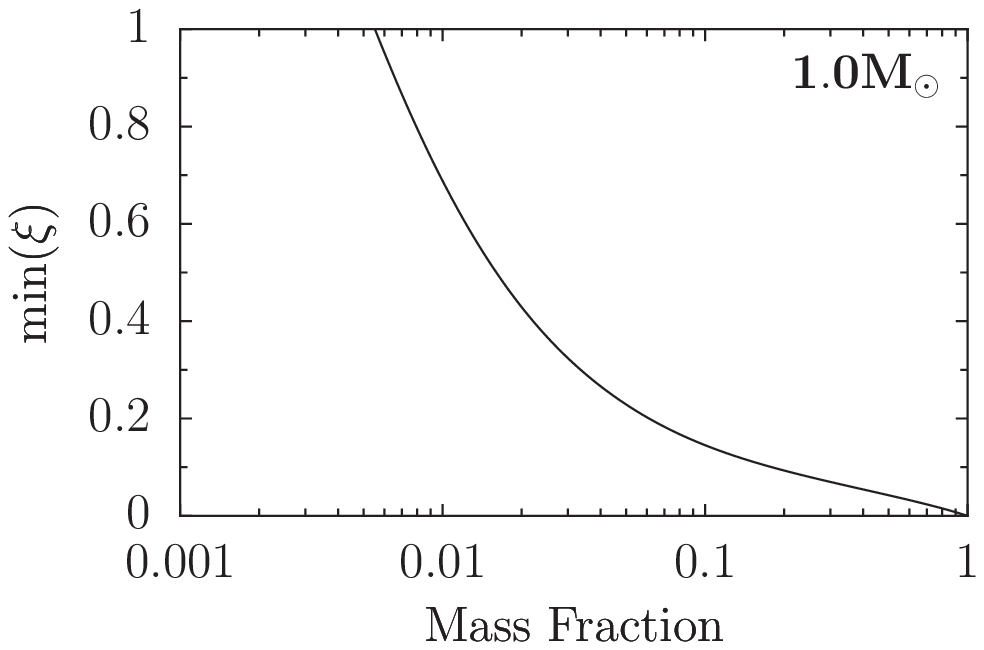}
	\caption{The compressibility of the unperturbed flow as a function of
	depth for the three simulation boxes ($0.775M_\odot$ - top left;
	$0.85M_\odot$ - top right; $1.0M_\odot$ - bottom). The horizontal axis
	gives the time averaged fraction of mass with
	compressibility greater than the vertical value.}
	\label{fig: compressibility}
\end{center}
\end{figure}

Because we are measuring the mass averaged dissipation, in order for the
perturbative treatment discussed above to be valid, we can only have 
$\xi\gtrsim1$ for a negligible fraction of the mass. In
figure \ref{fig: compressibility} we plot the time averaged fraction of mass
that resides in regions of the convective box which have $\xi$ greater than some
value. The value of the convective turnover time, necessary to evaluate $\xi$,
was calculated as $\tau_c\equiv\textrm{FWHM}(v_z)/\max(v_z)$, where
FWHM($v_z$) is the thickness of the layer over which $v_z>\max(v_z)/2$.

As we can see, in all cases, $\xi>1$ for less than 1\% of the mass.
This compression is concentrated near the top of the
box, where the density, and hence the dynamic viscosity, is small. So, even
though compressibility and shocks are important in determining the flow that
develops, no appreciable dissipation occurs in strongly compressible regions.
The situation is further improved by the fact that we apply a window in the
vertical direction that significantly reduces the importance, and completely
ignores part of the compressible driving region near the top of the box in
determining the effective viscosity (see section \ref{sec: Fourier windows}).

We partially treat the first source of compressibility discussed above by,
imposing $H_\rho=const$ in the continuity equation. However, this is not valid
for most astrophysically interesting convective zones. In fact in all cases 
considered here the density scale height varies by a factor of a few between the
top and the bottom of the convective layer. In some sense, assuming
$H_\rho=const$ is not any better than assuming $H_\rho=\infty$. We argue that
ignoring the stratification from the continuity equation is a reasonable
approximation.

The simplest way to justify this is to repeat the evaluation of the viscosity
with different values of $H_\rho$ within the range encountered in the convective
layer of interest.

We can also gauge the importance of the stratification by
comparing $d\ln\rho/dz$ to $\partial\ln\delta v_z/\partial z$. 
We evaluate $d\ln\rho/dz$ directly, and estimate:
\begin{equation}
	\frac{d\ln\delta v_z}{dz}=\frac{1}{\left<
	\delta {v_z}^2\right>^\frac{1}{2}}
	\left<\left(
		\frac{\partial\delta v_z}{\partial z}\right)^2
	\right>^\frac{1}{2}=\frac{1}{\left<
	{v_z}^2\right>^\frac{1}{2}}
	\left<\left(
		\frac{\partial v_z}{\partial z}\right)^2
	\right>^\frac{1}{2}.
	\label{eq: v log deriv}
\end{equation}
The last expression comes from equation \ref{eq: v perturbation}, and is correct
when the last row of $\mathbf{A}(t)$ contains only a single non-zero entry.
Since those are the only cases we use, this expression is sufficient for us.

\begin{figure}
\begin{center}
	\includegraphics[width=0.49\textwidth]{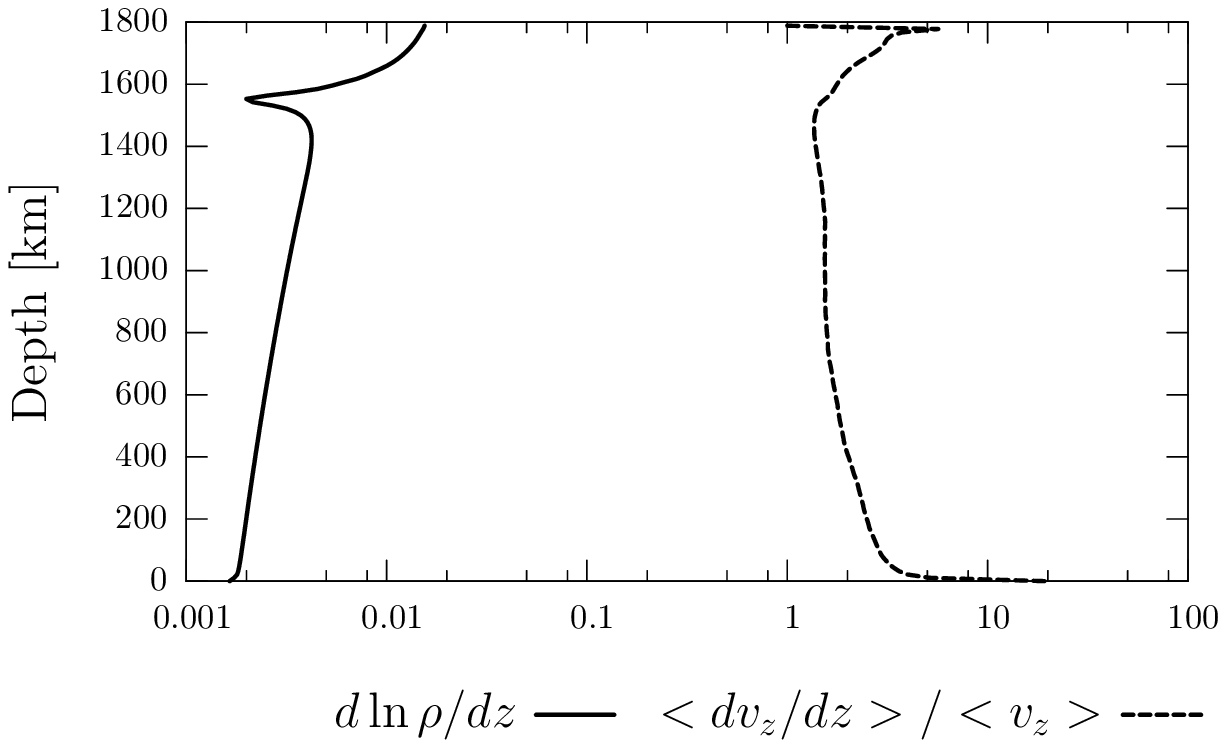}
	\includegraphics[width=0.49\textwidth]{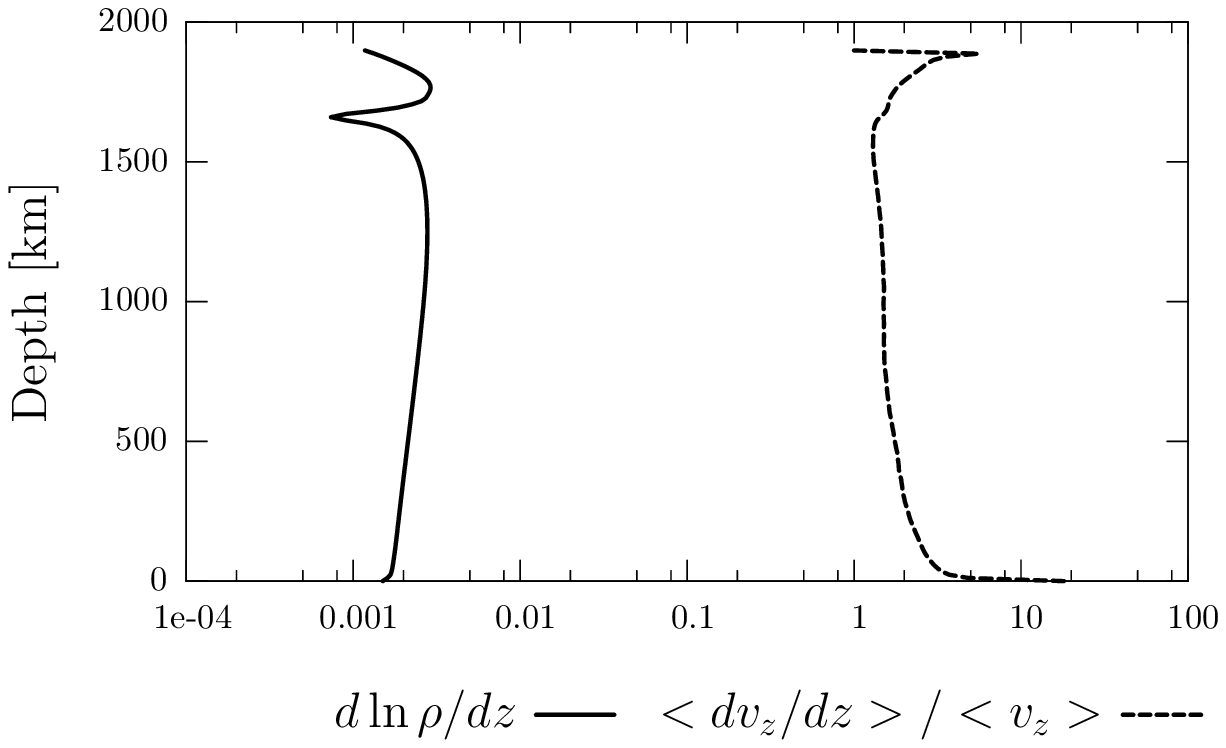}
	\includegraphics[width=0.49\textwidth]{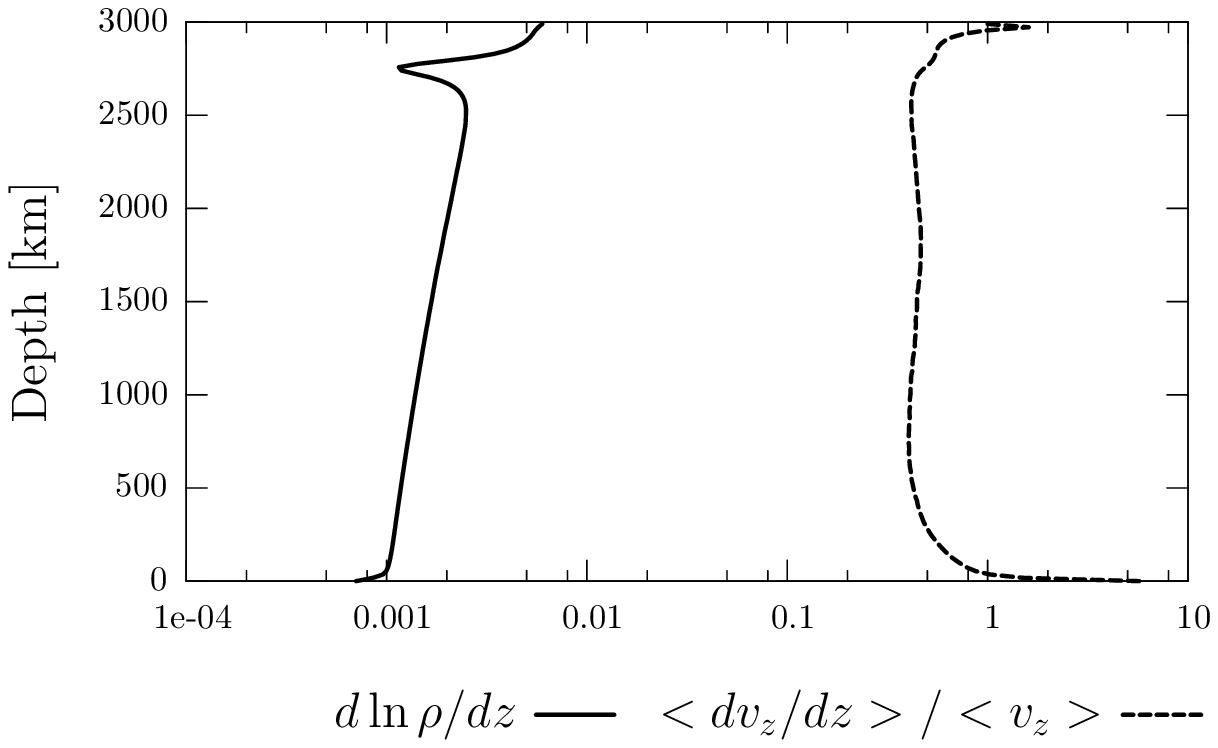}
	\caption{The logarithmic gradient of the density compared to the
	logarithmic gradient of $v_z$, estimated as explained in the text for
	the three simulation boxes we considered: $0.775 M_\odot$ - top left;
	$0.85 M_\odot$ - top right; $1.0M_\odot$ - bottom.}
	\label{fig: anelastic}
\end{center}
\end{figure}

In figure \ref{fig: anelastic} we compare $d\ln\rho/dz$ to $\partial\ln\delta
v_z/\partial z$ (estimated as in the above expressions). We see that the
logarithmic gradient of the density is approximately two orders of magnitude
smaller than the typical logarithmic velocity gradient and hence we are
justified in ignoring it.

\subsection{Radial and Time Fourier Transforms}
\label{sec: Fourier windows}
Using discrete Fourier transforms to represent a data set, forces the
assumption that the data is periodic in all dimensions. While this holds for 
each horizontal slice, it is violated for the radial and time dimensions.
Ignoring the problem leads to artificially introducing spectral power
at the highest frequencies because of the jumps at the boundaries. To avoid
this, we need a special way to deal with the non-periodic directions. 

The usual solution is to window the data so that it goes smoothly to zero at the
edges of the domain. This has the effect that it makes the values near the
center of the domain relatively more important than those near the
boundaries. Incidentally, this is exactly what we would like in the radial
direction, since the flow near the top and bottom is affected by the
artificial boundary conditions and is not representative of the actual flow that
would occur in a star.

Further, as discussed in section \ref{sec: compressibility} we expect that the
compressibility of the flow that we neglect might be significant in the upper
end of the box, where the density is small and the flow is supersonic. So making
this region's contribution to the overall dissipation small is exactly what we
would like. In fact, in the radial direction we go a step further and limit the
window to completely exclude some part of the box near the top and bottom
boundaries (see equations (\ref{eq: welch window}) and (\ref{eq: bartlett
window})). 

In the time direction, as long as the time interval we have
simulated is ``representative'' of the actual convection that occurs in a star,
weighing the center of the interval more than the edges should not be a problem.

To confirm that the chosen window function is not affecting our final result,
we derive the effective viscosity coefficient using two common windows:
\begin{eqnarray}
	\textrm{Welch} & : &
	\mathcal{N}\left[1-\left(\frac{2t-T}{T}\right)^2\right]
	\textrm{max}\left[0,1-\left(\frac{2z-L_z}{\alpha L_z}
	\right)^2\right]
	 \label{eq: welch window},\\
	\textrm{Bartlett} & : & 
	\mathcal{N}\left[1-\left|\frac{2t-T}{T}\right|\right]
	\textrm{max}\left[0,1-\left|\frac{2z-L_z}
	{\alpha L_z}\right|\right],
	\label{eq: bartlett window}
\end{eqnarray}
where $\mathcal{N}$ is a normalization factor numerically equal to the inverse
of the average  of the squares of the window function at all the grid points,
and $\alpha$ is a parameter determining what fraction of the radial span of the
box we include in the analysis, that is we exclude $(1-\alpha)$ fraction of
the linear size of the box, half from the top and half from the bottom.

\section{The Stellar Models}

The three models used, represent the top 7--9 pressure scale heights of the
convective zones of the present Sun, a 0.775 $M_\odot$ and a 0.85 $M_\odot$
stars. Table \ref{tbl: yale stars} shows the position of each model in
the $\log\,g-\log\,T_{\rm eff}$ plane. The full details of the numerical scheme
and the properties of the solar simulation are discussed in
\citet{Robinson_et_al_03}. For a comparison between the models used here and the
work of other groups, as well as observations see \citet{Kupka_05} and
\citet{Hillebrandt_Kupka_08}. Here we present very briefly only the most
important aspects of the models.

The simulation boxes have periodic side walls and impenetrable top and bottom
surfaces with a constant energy flux fed into the base and a perfectly
conducting top boundary. The imposed flux was computed from a corresponding 1D
stellar model with the chosen mass and age, thus  was not arbitrary, but the
correct amount of energy flux the computational domain should transport outward
in the particular star. The initial conditions of the 3D simulations were also
derived from the same 1D stellar models used to calculate the required flux.

\begin{table}
\begin{center}
\caption{The physical characteristics of the three simulations used to
derive effective viscosities. The units of $T_{\rm
eff}$ are K and the units of $g$ are $\textrm{cm}\,\textrm{s}^{-2}$.}
\label{tbl: yale stars}
\begin{tabular}{l|c|c|c}
Model Mass & $1\,M_\odot$ & $0.85\,M_\odot$ & $0.775\,M_\odot$\\
\hline
Age (Gyr) & $4.55$ & $7$ & $2$ \\
$\log\,T_{\rm eff}$ & $3.761$ & $3.685$ & $3.708$\\
$\log\,g $ & $4.44$ & $4.592$ & $4.592$ \\
Size (Mm) ($L_x\times L_y\times L_z$) & ${5.4}^2 \times 2.8$ & 
		${2.7}^2 \times 1.8$ & ${2.9}^2 \times 1.9$\\
Grid ($N_x\times N_y\times N_z$) & $114^2 \times 170$ & $115^2 \times 170$ &
		$115^2 \times 170$\\
${\rm R/R_\odot}$ & $1.0$ & $0.737$ & $0.772$
\end{tabular}
\end{center}
\end{table}

\subsection{Starting Models and Input Physics}
The 1D stellar models used to initialize each run were computed with the YREC
stellar evolution code \citep{Guenther_et_al_92}. They were calibrated to the
Sun and evolved from the ZAMS. Both the 1D and 3D codes use
the same realistic physics as described by \citet{Guenther_Demarque_97}, most
notably the \citet{Alexander_Ferguson_94} opacities at low temperatures, the
OPAL opacities and equation of state \citep{Iglesias_Rogers_96}, hydrogen and
helium ionization and helium and heavy element diffusion. 

Some details of the three models are given in Table \ref{tbl: yale stars}. The
fractional radius is given as ${\rm R/R_{\odot}}$, where
$R$ is radius of the stellar body and ${\rm R_\odot}$ is 
the radius of the Sun. Both are defined at the point where 
$T=T_{\rm eff}$. The surface gravity and effective temperature are in c.g.s.
units.

\subsection{Box dimensions}
The horizontal dimensions  of each computational box (column 5 in Table 1) were 
estimated by assuming that the granule size 
will scale roughly inversely with $g$. 
The final column gives the number of grid points in the two horizontal 
and vertical directions in the square based box.

\section{Results}
As discussed in section \ref{sec: anisotropic_visc} the anisotropic viscosity
can be parametrized by five independent components. Assuming equation (\ref{eq:
nu form}), we
evaluate those components using the two window functions of 
equations (\ref{eq: welch window}) and (\ref{eq: bartlett window})
each with two different values of $\alpha$: 0.8 and 0.9. In addition we use two
values for the density scale height in each case: $H_\rho=\infty$ and the volume
average density scale height.

The reason for using the volume averaged value of $H_\rho$ instead of the mass
averaged is that, this way, relatively more weight is given to the less dense
top regions where the density scale height is small, resulting in a larger range
between the two cases we consider.

\begin{figure}
\begin{center}
	\includegraphics[width=0.49\textwidth]{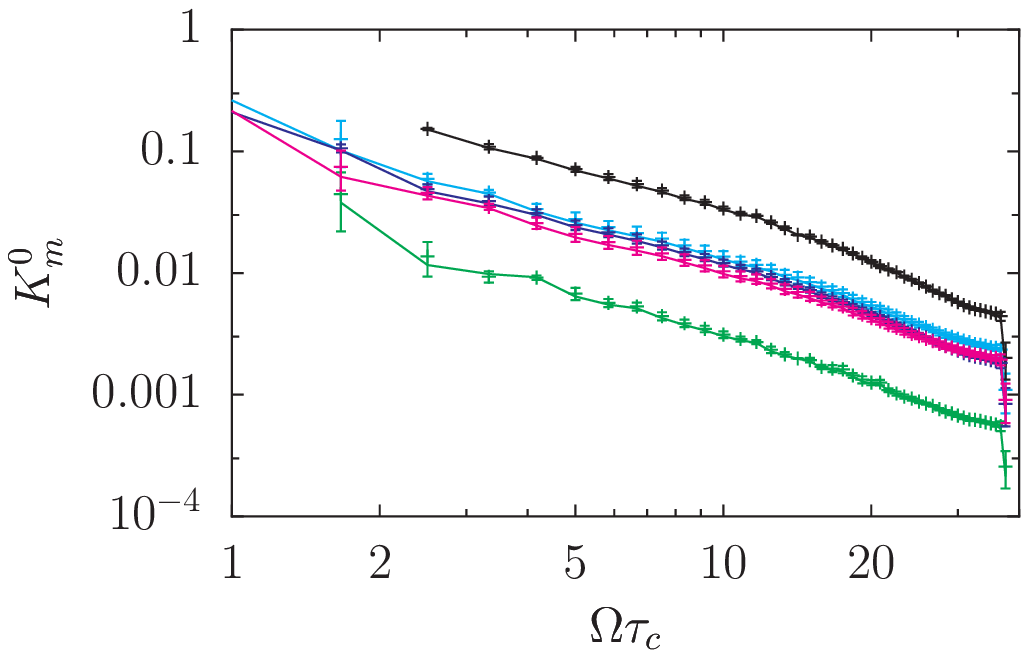}
	\includegraphics[width=0.49\textwidth]{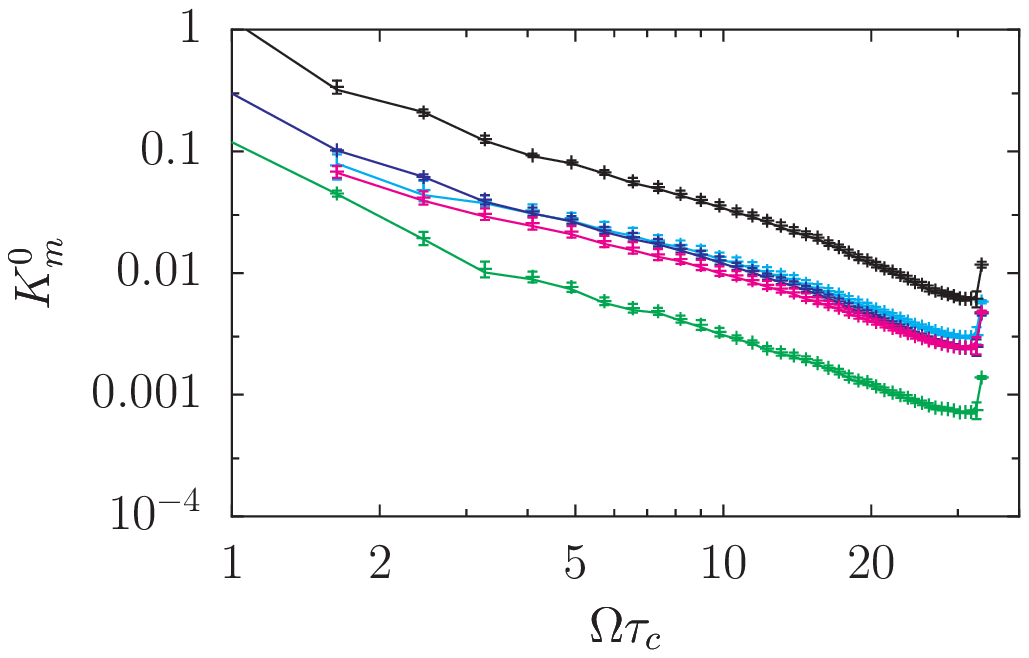}
	\includegraphics[width=0.49\textwidth]{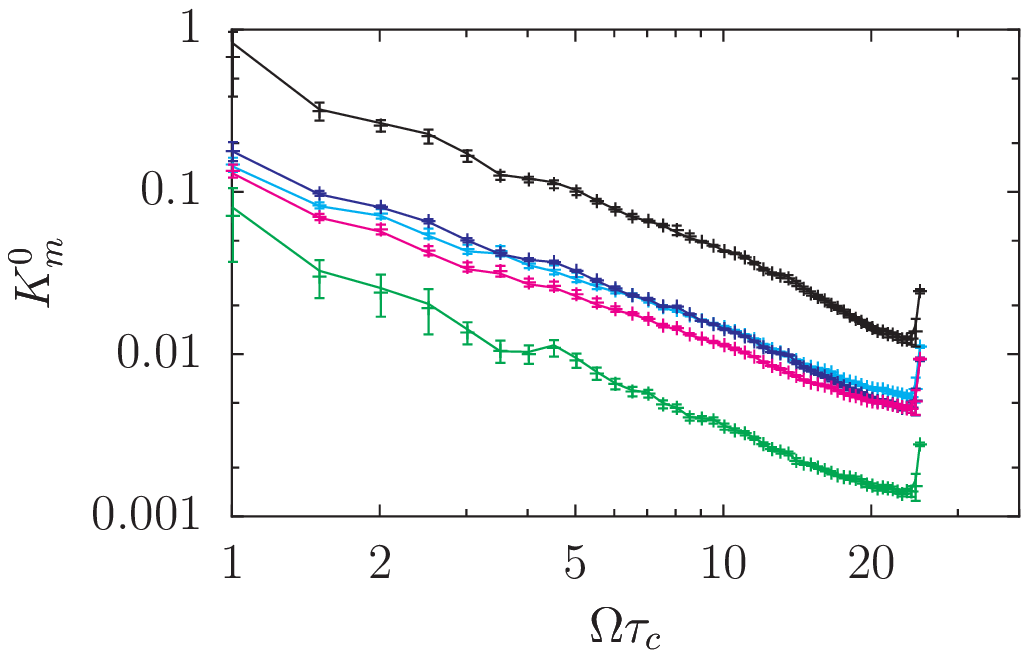}
	\includegraphics[width=0.7\textwidth]{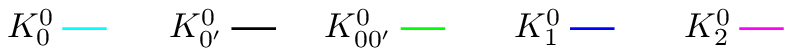}
	\caption{The five viscosity components, scaled
	by $\left<v^2\right>^{1/2}H_p$ for the three stellar models 
	($0.775M_\odot$ - top left; $0.85M_\odot$ - top right; 
	$1M_\odot$ - bottom). The
	lines correspond to the average of all curves representing the same
	component calculated using $\alpha=0.9$, Welch window (Eq.
	\ref{eq: welch window}) and the volume averaged density scale height.
	The error bars correspond to the spread found among all the curves
	corresponding to the viscosity component with different windows, values
	of alpha and density scale heights.}
	\label{fig: main result}
\end{center}
\end{figure}

The frequency dependence of the five viscosity coefficients 
($K_0$, $K_{0'}$, $K_{00'}$, $K_1$, and $K_2$) is
presented in figure \ref{fig: main result} the curves correspond to a Welch
window (Eq. \ref{eq: welch window}) with $\alpha=0.9$ and volume averaged
density scale height and the error bars show the span among all the cases for
which we evaluated the effective viscosity.

The fact that the error bars in figure \ref{fig: main result} are small shows
that indeed the choice of the window function is not important and that ignoring
the depth dependence of the density scale height, and in fact the stratification
altogether in the continuity equation is a valid approximation.

We see that the same qualitative characteristics hold for the estimated
dissipation in all 3 of our simulation boxes: the $K_{0'}$ component 
is always approximately four times larger than the $K_0$, $K_1$ and $K_2$
components, which are in turn roughly four to five times larger than the
$K_{00'}$ component and the scaling is approximately the same for all
components, close to the linear scaling proposed by Zahn. 

Quantitatively, the effective viscosity we calculate can be written as:
\begin{equation}
	K_m=\mathcal{K}_m^0 \rho\left<v^2\right>^{1/2} H_p
	\left(\frac{T}{\tau_c}\right)^\lambda, m\in\left\{0, 0', 00', 1,
	2\right\}
	\label{eq: effective visc}
\end{equation}
where the parameters $\mathcal{K}_m$ and $\lambda$ take the values:
\begin{equation}
	\begin{array}{r@{=}l}
	\lambda 		& 	1.2	\pm	0.3,\\
	\mathcal{K}_0		&	0.022	\pm	0.003,\\
	\mathcal{K}_{0'}	&	0.080	\pm	0.01,\\
	\mathcal{K}_{00'}	&	0.0046	\pm	0.0008,\\
	\mathcal{K}_1		&	0.024	\pm	0.003,\\
	\mathcal{K}_2		&	0.018	\pm	0.003,\\
	\end{array}
	\label{eq: K-dwarf results}
\end{equation}
with the errors corresponding to the range of values encountered for different
windows, values of $\alpha$, density scale heights and stellar models. The above
values were
derived by performing a least squares fit of equation \ref{eq: effective visc}
to the calculated curves for $T<0.5\tau_c$. The reason for restricting the fit
to short periods is that at long periods we do not expect the perturbative
calculation used in this work to be applicable and hence the derived slope is an
artifact of the model rather than having any physical significance. 

Also there is no appreciable difference between the models of the different
stars. The spread in the dimensionless effective viscosities for the three
models is not much larger than the error bars at all frequencies, except the
high end tails, where the effects of the finite resolution and time sampling
become important. This suggests that at least for the range of conditions
encountered in the convective zones of low mass stars the dissipation
efficiency is not strongly dependent on the details of the convective flow.

\subsection{Anisotropy}
The above splitting of the viscosity in five components was done in order to
allow for anisotropic dissipation. It is interesting to see how anisotropic the
derived effective viscosity really is. The general isotropic case has only two
viscosity components: a bulk viscosity ($\zeta$) and a shear viscosity ($\eta$)
(c.f. \citet{Landau_Liefshitz}). In terms of those the isotropic $K_{ijmn}$
tensor is:
\begin{equation}
	K_{ijmn}=\eta\left(\delta_{im}\delta_{jn} + \delta_{in}\delta_{jm} -
	\frac{2}{3}\delta_{ij}\delta_{mn}\right)  + \zeta\delta_{ij}\delta_{mn}
\end{equation}
From this it can be seen that the five components of the viscosity we calculate
must obey the relations:
\begin{equation}
	\begin{array}{c}
	K_1=K_2=\eta\\
	K_0=K_{00'}+K_1\\
	K_{0'}=K_0+K_1.\\
	\end{array}
	\label{eq: isotropic relations}
\end{equation}
We see that the first two of these are clearly satisfied by the viscosity
coefficients of equation \ref{eq: K-dwarf results} to within the quoted
uncertainties. The degree to which the last equation is not satisfied is:
\begin{equation}
	\frac{K_{0'}}{K_0+K_1}-1=0.74\pm0.25.
	\label{eq: anisotropy}
\end{equation}

Considering the fact that the flows in our simulation boxes are not exactly like
those inside stars, and the loosely estimated errors in equation \ref{eq:
K-dwarf results} we can conclude that the effective viscosity we find is only
mildly anisotropic, and it is perhaps reasonable to approximate it as completely
isotropic bulk and shear viscosities with the following values:
\begin{equation}
	\begin{array}{rl}
		\eta=&0.020\pm0.003\\
		\zeta\in&(0.018,0.056).
	\end{array}
\end{equation}

The reason for $\zeta$ not being well determined by the viscosity coefficients
(\ref{eq: K-dwarf results}) is that those coefficients do not exactly correspond
to an isotropic viscosity (see equation (\ref{eq: anisotropy})).

\section{Conclusion}
We have extended the analysis of \citet{Penev_Sasselov_Robinson_Demarque_07} to
calculate effective viscosities in the surface convective zones of three main
sequence stars: $0.775 M_\odot$, $0.85 M_\odot$ and the present day Sun. We have
also modified the calculation to properly account for all normalization factors.

The effective viscosity we find (given by equations (\ref{eq: effective visc})
and (\ref{eq: K-dwarf results})) scales linearly with the period of the external
perturbation, with the shear viscosity being smaller than the linear
scaling proposed by \citet{Zahn_66, Zahn_89} by a factor of about ten, but in
addition there is a significant bulk viscosity, which is assumed zero in Zahn's
prescription.

This factor in practice does not have a dramatic effect on the tidal
circularization period, which scales as
$K^{3/16}$ \citep{Zahn_66, Zahn_89}. So, assuming that the above effective
viscosity is correct in the range of periods applicable to stellar binary
orbits, and that the saturation period is $2\tau$ as assumed by \citet{Zahn_66,
Zahn_89}, the circularization cut-off period based on our viscosity would be
within about 30\% of the prediction with Zahn's scaling.

The important difference between this effective viscosity and equation (\ref{eq:
Zahn viscosity}) is the presence of a significant bulk viscosity, the
possibility that the effective viscosity is not isotropic, and
that the linear scaling should apply only for a limited range of frequencies.

The applicability of this result is limited by two factors: the range of
applicability of the perturbative expansion (Eq. \ref{eq: e dot}) and the
limits of the numerical simulations.

The external shear velocities are assumed, by the perturbative expansion, to
be small compared to the typical convective flow, and the period of the external
shear should be neither too long nor too short. 

On one hand the limited spatial resolution of the numerical simulations means
that only
sufficiently large turbulent eddies are captured, which implies that our results
do not apply to external forcing with very short period, for which the
dissipation may be dominated by eddies that are too small to be reliably
simulated. However, sufficiently short periods fall within the inertial
subrange where Kolmogorov scaling holds and in that case the same perturbational
calculation predicts quadratic scaling of the effective viscosity with period
\citep{Goodman_Oh_97}. 

On the other hand, the perturbative expansion we use, assumes that the
perturbation period ($T$) is small compared to the turnover time ($\tau$) of
the largest local eddies. In particular, we expect that the effective viscosity
should reach a maximum value for some perturbation period on the order of
$\tau$, and remain the same for all longer periods. This saturation cannot be
captured by our perturbative approach since it is due to the neglected higher
order terms.

\citet{Penev_Barranco_Sasselov_08b} used a spectral, anelastic, ideal gas
convective box, which includes the external forcing as part of the equations of
motion, to find the effective viscosity directly without a perturbative
treatment. They confirm that the slope of the perturbative viscosity is
consistent with the directly calculated values in the range of its
applicability, although for the $x-z$ component they observe a period
independent offset between the perturbative and direct viscosity which acts to
increase the anisotropy. 
They also find linear scaling of the effective viscosity with period
that saturates for $T>2\tau_c$. The magnitude of the effective viscosity they
find based on the perturbative calculation described above is approximately a
factor of two larger than the results presented in this paper. However, this is
due to the fact that the \citet{Penev_Barranco_Sasselov_08b} convective zone has
a mixing length parameter of about 3: double the value usually assumed for the
Sun and appropriate for the simulations used above.

We would like to thank the anonymous referee for detailed discussion of
the parametrization of the viscosity which improved the paper considerably. We
would also like to acknowledge much helpful advice that generally improved the
quality of this work from Dr. Jeremy Goodman.

\bibliography{convective_turbulence}
\bibliographystyle{apj}
\end{document}